\pdfoutput=1
\documentclass[10pt,twocolumn]{article}
\usepackage[clean,nolineno]{loc-style}

\newcommand{\figdose}{\autoref{fig:dose_curve}(a)}
\newcommand{\figtemporal}{\autoref{fig:dose_curve}(b)}
\newcommand{\figdevvar}{\autoref{fig:dose_curve}(c)}

\DeclareRobustCommand{\suppref}[1]{ESI\textsuperscript{\dag} Section~#1}

\begin{document}

\twocolumn[
\begin{@twocolumnfalse}
\vspace*{-4mm}
{\sffamily\small\color{gray}Paper}\par\vspace{4pt}
{\sffamily\bfseries\LARGE Multiplexed Label-Free Cytokine Sensing Using a 128$\times$128 CMOS \isfet{} Array with \flexdym{} Microfluidic Partitioning\par}
\vspace{8pt}
{\sffamily\large Utku Noyan,\textsuperscript{a,b} Sahil Shah\textsuperscript{*a} and Pamela Abshire\textsuperscript{*a,b,c}\par}
\vspace{8pt}
\locabstractbox{%
Cytokine-mediated inflammation disorders such as sepsis and cytokine release syndrome demand bedside
multiplex panels that resolve pro- and anti-inflammatory dynamics in minutes, yet label-free FET cytokine sensors remain limited to few devices, single analytes, and PDMS packaging that sorbs small molecules and leaches oligomers. We present a CMOS platform pairing a $128 \times 128$ ion-sensitive field-effect transistor (ISFET) array (SkyWater \SI{130}{nm}, SiN gate) with a FlexDym $4 \times 1$ micro-well overlay that partitions the membrane into four chemically independent regions ($\approx \pxwell$ pixels each, $\sqrt{N}$ averaging). Two wells carry capture antibodies (anti-IL-6, anti-IL-10), one is a non-specific reference, and one tests specificity; the transduced signal is defined as the differential current relative to the reference signal ($\dIds(W_n)-\dIds(W_3)$).  Single-analyte dose-response measurements spanning \SI{100}{fM}--\SI{10}{nM} successfully resolve both IL-6 and IL-10 above the non-specific detection floor, including at healthy-serum baseline concentrations. The array shows a per-pixel pH sensitivity of \SI{1.13}{\micro\ampere/\text{pH}} ($\text{CV} \sim 14\%$). Across two chips, device-to-device variation is dominated by transconductance spread ($g_m$ $\text{CV} \approx \devraw$); a one-time per-device $g_m$ calibration reduces analyte-measurement variance to $\approx \devcalib$. We further validate the surface functionalization, characterize device-to-device variation, and demonstrate the FlexDym integration workflow. Simultaneous multiplexed cytokine measurements resolve IL-6 (W1) and IL-10 (W2) against the chemistry-only reference (W3). At the same time, cross-specificity characterization is assessed by applying IL-6 to an anti-IL-10 well (W4), establishing a scalable label-free ISFET architecture for multiplexed cytokine sensing.}
\end{@twocolumnfalse}
]

{\renewcommand{\thefootnote}{}%
\footnotetext{\footnotesize
\textsuperscript{a}\,Department of Electrical and Computer Engineering, University of Maryland, College Park, MD 20742, USA.\\
\textsuperscript{b}\,Institute for Systems Research, University of Maryland, College Park, MD 20742, USA.\\
\textsuperscript{c}\,Department of Electrical and Computer Engineering, Stony Brook University, Stony Brook, NY 11790, USA.\\
\textsuperscript{*}\,Corresponding authors. E-mail: sshah389@umd.edu; pamela.abshire@stonybrook.edu\\
\textsuperscript{\dag}\,Electronic supplementary information (ESI) available: surface-functionalization protocol, per-pixel sensitivity and pH reversibility, noise and drift analysis, temporal kinetics, masked-well dose maps and clinical concentration mapping.}}

\ifloclineno\linenumbers\fi

\section{Introduction}
\label{sec:intro}
Cytokines are short-lived, low-abundance signaling proteins whose temporal trajectories carry decisive diagnostic information about many inflammatory disorders including sepsis, acute respiratory distress syndrome, cytokine release syndrome, and chimeric antigen receptor T-cell (CAR-T) toxicity\cite{FajgenbaumJune2020,Pierrakos2010,Lee2014CRS}. Interleukin-6 (IL-6) is a canonical pro-inflammatory mediator; serum levels rise from \textless\SI{10}{pg/mL} in healthy adults to \textgreater\SI{1}{ng/mL} during sepsis or cytokine release syndrome (CRS), and the timing of that rise governs therapeutic windows for tocilizumab and other IL-6-axis interventions\cite{Tanaka2014IL6}. Interleukin-10 (IL-10) is the principal counter-regulatory anti-inflammatory cytokine; the IL-6/IL-10 ratio outperforms either marker alone for prognostic stratification in septic shock\cite{Moore2001,Frieler2025IL6IL10ratio}. A fast, simultaneous IL-6/IL-10 readout is therefore the smallest meaningful inflammation panel.

ELISA delivers pg/mL sensitivity but requires multiple wash steps and \SIrange{3}{5}{\hour} of bench time per analyte\cite{Engvall1971ELISA}. Multiplex bead assays (Luminex) consolidate panels but demand a flow cytometer and \SIrange{3}{4}{\hour} workflows\cite{Vignali2000Luminex}. Lateral-flow strips are bedside-deployable but typically resolve only $\geq$\SI{10}{pg/mL} and are endpoint assays, missing the kinetics that matter most\cite{Posthuma2007Luminex}. Optical multiplex platforms based on localized surface plasmon resonance approach pg/mL sensitivity\cite{Chen2017LSPRcytokines} but retain optical instrumentation and fluidic complexity that resist handheld miniaturization.

Field-effect transistor biosensors offer a CMOS-compatible, label-free, optics-free alternative that converts charge-on-binding into a threshold-voltage shift\cite{Bergveld1970, Bergveld2003, Stern2007Debye}. Rani \etal{} demonstrated sub-picomolar IL-2/IL-4 detection on \emph{four}-pixel silicon-nanowire \isfet{} arrays\cite{Rani2020}; Hu \etal{} reported a \SI{2.1}{pg/mL} IL-6 LOD on a single silicon-nanowire FET using a DNA aptamer\cite{Hu2023IL6}. Both share three structural limits: (i)~the active sensor count per chip is small (\textless10), so cluster-level averaging and yield-aware referencing are unavailable; (ii)~the platforms target a single analyte per device, so multiplexing requires fluidic addressing rather than spatial chemistry partitioning; and (iii)~the supporting fluidic packaging is bonded glass or PDMS, both of which carry biocompatibility caveats for hydrophobic-analyte assays\cite{Toepke2006, Regehr2009}.

Parallel CMOS \isfet{} work has pushed pixel counts into the millions for sequencing\cite{Rothberg2011IonTorrent} and demonstrated drift-compensated detection at \num{12800}-pixel and \num{1}M-pixel scales\cite{Miscourides2018, Moser2018Scalable}; aptamer-modified CMOS \isfet{} arrays reached \SI{3.27}{ng/mL} troponin~I\cite{Ryazantsev2024}, and a \SI{0.35}{\micro\meter}-CMOS biosensor reported \SI{45}{pM} IL-6 with ELISA correlation\cite{Yang2018PreClinical}. None has combined a dense statistics-rich CMOS array, a low-adsorption microfluidic partition, and a multiplexed cytokine panel with built-in reference channels.

This paper presents that combination. We integrate a SkyWater \SI{130}{nm} PMOS \isfet{} pixel array ($128\times128=\num{16384}$ pixels, \siN{} gate) with a \flexdym{} $4\times1$ micro-well overlay. \flexdym{} is a styrene-ethylene-butylene-styrene (SEBS) thermoplastic elastomer offering ISO~10993 biocompatibility, reversible bonding, low small-molecule sorption, and no oligomer leaching---the four limitations of PDMS that have constrained quantitative protein sensing\cite{Lachaux2017FlexDym, Roy2011SEBS, EdenFlexDym}. The four wells are functionalized as W1: anti-IL-6, W2: anti-IL-10, W3: chemistry-only (APTES$+$GA) reference, and W4: specificity well, and the differential signal $\dIds(W_n)-\dIds(W_3)$, $n=1,2$ rejects confounding signals related to pH-strength, temperature, and ionic drift.

\subsection{Contributions}
\label{subsec:contrib}
\begin{enumerate}[leftmargin=*,nosep]
  \item A CMOS \isfet{} pixel-array platform for multiplexed label-free cytokine sensing that scales array statistics to $\approx\pxwell{}$ pixels per well, versus prior FET cytokine sensors with $\leq$4 active sensors\cite{Rani2020,Hu2023IL6}.
  \item The first integration, to our knowledge, of a \flexdym{} (SEBS) microfluidic overlay with a CMOS \isfet{} array, eliminating the sorption/leaching pathologies of PDMS\cite{Toepke2006,Regehr2009,Lachaux2017FlexDym}.  
  \item Measured single-analyte IL-6 \emph{and} IL-10 dose-responses with
  whole-well $\sqrt{N}$ averaging, yielding $3\sigma$ LODs that reach the
  healthy-serum baseline for IL-10 and the clinically-elevated range for IL-6,  within a four-well differential architecture with chemistry-only and
  reference channels.
  \item A two-chip device-to-device characterization that quantifies transconductance-driven measurement variance and shows that a per-device $g_m$ calibration ($\dVth=\dIds/g_m$) reduces it from $\approx\devraw{}$ to $\approx\devcalib{}$.
  \item A reproducible recipe transfer from prior phosphatidylserine detection\cite{NoyanPS2024}, demonstrating a transferable surface functionalization protocol for amine-coupled probes.
\end{enumerate}

\section{Background and Related Work}
\label{sec:related}

\subsection{Cytokines as Multiplex Analytes}
\label{subsec:cytobio}
The cytokine network is multivariate: it encompasses the relative dynamics of pro- and anti-inflammatory mediators that distinguish resolving inflammation from progressing sepsis\cite{Galluzzi2018, FajgenbaumJune2020}. The IL-6/IL-10 pair is the canonical minimal panel, and their ratio improves prognostic accuracy over either marker alone\cite{Frieler2025IL6IL10ratio}. Both are small monomeric proteins ($\sim$\SI{21}{kDa} IL-6, $\sim$\SI{18.6}{kDa} IL-10) with well-characterized capture antibodies, making them tractable for immobilization-based field-effect transduction.

\subsection{FET-Based Cytokine Sensors}
\label{subsec:fet_cyto}
FET-based cytokine sensors have demonstrated highly sensitive, label-free detection across a range of transducer architectures. Rani \etal{}\cite{Rani2020} reported sub-picomolar detection of IL-2 and IL-4 using nano\isfet{} arrays, establishing the feasibility of real-time cytokine monitoring with field-effect transduction. Hu \etal{}\cite{Hu2023IL6} achieved an IL-6 limit of detection of \SI{2.1}{pg/mL} using an aptamer-functionalized silicon nanowire FET, highlighting the sensitivity gains attainable with compact recognition elements. Extending field-effect biosensing to integrated platforms, Ryazantsev \etal{}\cite{Ryazantsev2024} demonstrated aptamer-based troponin-I detection on a CMOS \isfet{} array, while Yang \etal{}\cite{Yang2018PreClinical} reported \SI{45}{pM} IL-6 detection using a \SI{0.35}{\micro\meter} CMOS biosensor and validated the results against ELISA measurements.

\autoref{tab:cytolit} places these studies within the broader landscape of label-free cytokine biosensors. Individual FET transducers have achieved fM--pg/mL sensitivity for single cytokines\cite{Lit13_SiNW_IL2_Sandwich,Lit17_MoS2_TNF,Lit08_PEGAu_IL8}, while integrated microelectrode platforms have demonstrated multiplex cytokine detection on a common substrate\cite{Lit09_MEA_IL1IL10}. Collectively, these efforts illustrate the trade-offs between sensitivity, multiplexing, integration, and scalability. The present work builds on these advances by combining a dense CMOS \isfet{} pixel array fabricated in a standard CMOS process with spatial microfluidic partitioning, differential referencing, and array-level statistical averaging, leveraging the scalability of semiconductor manufacturing to integrate thousands of sensors for multiplex cytokine sensing on a single chip.
\begin{table*}[!t]
  \centering
  \caption{Survey of label-free cytokine biosensors by transduction modality, with limit of detection, dynamic range, response time, CMOS integration, and multiplex capability.}
  \label{tab:cytolit}
  \renewcommand{\arraystretch}{1.25}
  \setlength{\tabcolsep}{4pt}
  \footnotesize
  \begin{tabularx}{\textwidth}{@{}p{1.8cm} p{1.9cm} >{\centering\arraybackslash}p{0.9cm} p{2.3cm} p{1.5cm} p{1.4cm} >{\centering\arraybackslash}p{0.9cm} X@{}}
    \toprule
    \textbf{Analyte} & \textbf{Transduction} & \textbf{CMOS} & \textbf{LoD} & \textbf{Range} & \textbf{Time} & \shortstack{\textbf{Multi-}\\\textbf{plex}} & \textbf{Platform notes / Ref.} \\
    \midrule
    IFN-$\gamma$, IL-10 & Impedance          & $-$ & 25, 46\,pg/mL   & 0.1--5, 0.1--2\,ng/mL & 30\,min incub.    & $+$ & Aerosol-jet-printed graphene films, dual-Ab\cite{Lit01_AerosolJetGraphene} \\
    IL-6                & Impedance          & $-$ & 0.33\,pg/mL     & 1--15\,pg/mL          & 30\,min incub.    & $-$ & Au $+$ polypyrrole NP aptasensor\cite{Lit03_PolypyrroleIL6} \\
    IL-6                & Electrochemical    & $-$ & 10\,pM          & ---                   & 1\,h incub.       & $-$ & PEDOT:PSS, IL-6 receptor\cite{Lit07_PEDOTIL6} \\
    IL-1, IL-10         & Electrochem. \& Imp. & $-$ & 0.3, 0.7\,pg/mL & 1--15\,pg/mL       & 30\,min incub.    & $+$ & Fully-integrated microelectrode array\cite{Lit09_MEA_IL1IL10} \\
    IL-6                & FET (CNT)          & $-$ & 1.37\,pg/mL     & ---                   & 15\,min incub.    & $-$ & CNT FET, anti-IL-6 receptor\cite{Lit11_CNT_IL6} \\
    IL-6                & FET (Si NW)        & $-$ & 2.1\,pg/mL      & ---                   & ---               & $-$ & Si-nanowire FET, DNA aptamer; single device\cite{Hu2023IL6} \\
    IL-2                & FET (Si NW)        & $-$ & 10--14\,fM      & 20\,fM--200\,pM       & 3.5\,h incub.     & $-$ & Si NW, sandwich $+$ DNA amplification\cite{Lit13_SiNW_IL2_Sandwich} \\
    TNF-$\alpha$        & FET (MoS$_2$)      & $-$ & 60\,fM          & 6\,fM--6\,pM          & 2\,h incub.       & $-$ & Few-layer MoS$_2$, anti-TNF Ab\cite{Lit17_MoS2_TNF} \\
    IL-2, IL-4          & FET (nano\isfet{}) & $-$ & 3--5\,fM        & 3\,fM--1\,$\mu$M      & 15\,min real-time & $+$ & nano\isfet{}s in T-cell culture\cite{Rani2020} \\
    \midrule
    \textbf{This work}\newline IL-6, IL-10 & \isfet{} array & $+$ &
      \begin{tabular}[t]{@{}ll@{}} IL-6:  & \LODilsixMass{} \\ IL-10: & \LODiltenMass{} \end{tabular} &
      100\,fM--10\,nM & 20--30\,min, real-time & $+$ &
      $128\times128$ \flexdym{} $4\times1$ micro-wells; differential referencing \\
    \bottomrule
  \end{tabularx}
\end{table*}
\subsection{CMOS \isfet{} Pixel Arrays}
\label{subsec:cmos_arrays}

Rothberg \etal{}\cite{Rothberg2011IonTorrent} demonstrated million-pixel CMOS \isfet{} arrays for semiconductor sequencing, establishing the scalability of dense ion-sensing architectures. Moser \etal{}\cite{Moser2018Scalable} extended this concept with automatic calibration schemes for large-scale nucleic-acid amplification assays, while Miscourides and Georgiou\cite{Miscourides2018} analyzed current- and voltage-mode readout strategies in a \num{12800}-pixel array. Collectively, these studies showed that pixel-to-pixel offset, drift, and 1/f noise can be mitigated through differential measurement and calibration, and that large sensor populations enable statistically robust signal extraction through spatial averaging. However, these platforms were developed primarily for nucleic-acid sensing and sequencing applications. The present work applies the same dense-array sensing paradigm to multiplexed protein detection, integrating a statistics-rich CMOS \isfet{} array with a microfluidic partitioning layer designed for spatially resolved cytokine assays.

\subsection{Microfluidic Partitioning Materials}
\label{subsec:fluidics}

PDMS has been the dominant material for microfluidic partitioning because of its optical transparency, ease of fabrication, and widespread adoption in biological assays\cite{Whitesides2006}. However, several material properties can complicate quantitative protein sensing: (i)~PDMS absorbs hydrophobic molecules, altering reagent concentrations and assay kinetics\cite{Toepke2006}; (ii)~it can release uncrosslinked oligomers that adsorb onto nearby surfaces and affect sensor stability\cite{Regehr2009}; and (iii)~permanent bonding typically requires oxygen-plasma activation, which may interfere with surface-functionalization workflows.

SEBS, commercialized as \flexdym{}, retains the elastomeric handling characteristics of PDMS while addressing several of these limitations\cite{EdenFlexDym}. Roy \etal{}\cite{Roy2011SEBS} introduced SEBS for rapid microfluidic prototyping, and Lachaux \etal{}\cite{Lachaux2017FlexDym} demonstrated rapid hot embossing, reversible bonding, and substantially reduced small-molecule absorption. Commercial \flexdym{} formulations are ISO~10993 and USP Class~VI compliant\cite{EdenFlexDym}, form room-temperature seals against \siN{} surfaces without plasma activation, and can be reversibly bonded with reported pressure tolerance approaching \SI{4}{bar}.

These characteristics make SEBS attractive for CMOS biosensor integration, where preservation of the sensing surface and compatibility with post-fabrication functionalization are critical. To our knowledge, no prior CMOS \isfet{} biosensor has employed SEBS as the microfluidic partitioning material. Bonding and cross-leak validation are presented in \suppref{S-II}.

\section{Platform Architecture}
\label{sec:arch}

This section describes the integrated sensing platform, comprising the CMOS \isfet{} array, the \flexdym{} microfluidic partitioning layer, and the differential sensing strategy used for multiplexed cytokine detection. The hardware architecture establishes spatially isolated sensing regions with thousands of ISFETs assigned to each assay well, while the differential measurement framework combines cluster-level averaging with reference-well subtraction to improve measurement robustness and suppress common-mode disturbances.
\begin{figure*}[!t]
  \centering
  \includegraphics[width=\textwidth]{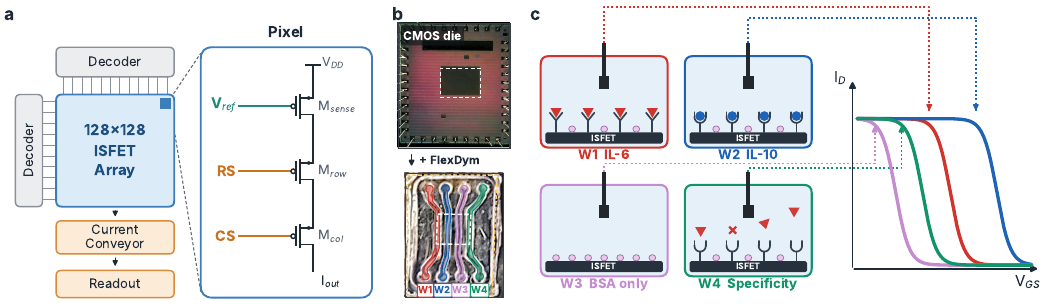}
\caption{System overview.
    \textbf{(a)}~Chip architecture and pixel schematic. The SkyWater \SI{130}{nm} chip is a
    $128\times128$ ISFET array addressed by row/column decoders and read out through a current conveyor (CCII). Each 3T pixel has a PMOS sensing transistor
    $M_{\mathrm{sense}}$, gate exposed to the electrolyte through a foundry-passivated $\text{Si}_3\text{N}_4$ membrane, with row- and column-select transistors
    $M_{\mathrm{row}}$, $M_{\mathrm{col}}$; \texttt{Pix\_ref} sets the electrolyte bias and the
    pixel current $I_{\mathrm{out}}$ is steered to the CCII readout.
    \textbf{(b)}~Optical micrograph of the packaged chip with the FlexDym $4\times1$ micro-well
    overlay bonded to the die; the four colored channels are the isolated wells aligned to the
    pixel clusters.
    \textbf{(c)}~Multiplex assay concept. The wells are functionalized as anti-IL-6 (W1; 5IL6,
    mouse), anti-IL-10 (W2; JES3-9D7-AF488, rat), a non-specific BSA-only reference (W3), and a
    specificity/off-target challenge (W4); antibody--antigen binding shifts the ISFET threshold
    voltage, giving the colour-matched $I_D$--$V_{GS}$ shifts, while the reference and
    off-target wells show negligible shift.}
  \label{fig:overview}
\end{figure*}

\subsection{CMOS \isfet{} Array}
\label{subsec:array}
The sensing chip is fabricated in the SkyWater \SI{130}{nm} CMOS process\cite{SkyWaterPDK} and contains a $128\times128$ array of PMOS \isfet{} pixels (\autoref{fig:overview}a) on a \SI{10}{\micro\meter} pitch, giving a $\SI{1.28}{\milli\meter}\times\SI{1.28}{\milli\meter}$ active area. Each pixel exposes a floating gate to the electrolyte through the foundry passivation stack terminated in \siN{}.

\textit{Pixel circuit.}~The 3T cell uses one PMOS sensing device $M_{\mathrm{sense}}$ (floating gate, electrolyte-exposed) in series with $M_{\mathrm{row}}$ and $M_{\mathrm{col}}$ access transistors; \texttt{Pix\_ref} is the reference electrode voltage which determines $V_{SG}$. When addressed, $M_{\mathrm{sense}}$ sources a drain current $I_{D}$ that depends on the surface potential $\psi_0$ at the \siN{}--electrolyte interface, hence on bound charge in the Debye layer.

\textit{Readout.}~On-chip, each addressed pixel drives a column-shared 
current conveyor that holds the column at a virtual reference (fixing $V_{DS}$) and copies $I_D$ off the array at unity current gain; the row/column decoders and the current conveyor are integrated on the CMOS die, producing one $128 \times 128$  current frame per raster. The primary metric is the drain current $I_D(t)$ 
at fixed bias; per-pixel $g_m$ from $I_D(V_{GS})$ sweeps converts $\dIds$
to an equivalent $\dVth$ for cross-device reporting (Section~\ref{subsec:res_dev}).

\subsection{\flexdym{} \texorpdfstring{$4\times1$}{4x1} Micro-Well Overlay}
\label{subsec:flexdym}
The \flexdym{} overlay (\autoref{fig:overview}b) is replica-molded from an SU-8 master and aligned so each well covers a known pixel cluster. The measured geometry is \SI{170}{\micro\meter}-wide wells separated by \SI{200}{\micro\meter} walls; at the \SI{10}{\micro\meter} pitch this maps to \num{17} pixel columns per well and \num{20} columns per wall ($4\times17+3\times20=128$ exactly). Each well therefore covers $17\times128\approx\pxwell{}$ pixels and provides $\sqrt{N}\approx\rootN{}\times$ noise averaging relative to a single pixel. Each well is a \SI{500}{\micro\meter}-high channel that runs across the full array from an inlet reservoir to an outlet reservoir; the solution dispensed into a well (\SI{50}{\micro L}) passes over the sensing pixels and collects in the outlet reservoir, so the channel volume above the array is only a small fraction of the loaded volume. After surface functionalization (Section~\ref{subsec:amine}), the \flexdym{} overlay is activated in O$_2$ plasma (\SI{150}{W}, \SI{30}{s}), placed on the functionalized die, and bonded on a hotplate at \SI{90}{\celsius} for \SI{30}{s} followed by \SI{120}{\celsius} for \SI{2.5}{min}\cite{EdenFlexDym}, and the bond is mechanically reversible. An $8\times1$ layout ($\sim$\num{1000} pixels per well) is supported in the same area for higher-multiplex panels. Bonding mechanics and cross-leak isolation are discussed in \suppref{S-II}.

\subsection{Differential Signal Model}
\label{subsec:signal}
The \isfet{} threshold voltage responds to bound-charge density $\sigma_{\mathrm{bound}}$ through the surface potential, screened over the Debye length\cite{Stern2007Debye},
\begin{equation}
  \dVth \propto \frac{\sigma_{\mathrm{bound}}}{C_{\mathrm{dl}}}\,\exp\!\left(-\frac{d}{\lD}\right),
  \label{eq:dVth}
\end{equation}
where $C_{\mathrm{dl}}$ is the double-layer capacitance and $d$ the charge--surface distance. In $1\times$ PBS, $\lD\approx\SI{0.7}{nm}$; in $0.01\times$ PBS, $\lD\approx\SI{7.4}{nm}$, comparable to the IgG--cytokine height. All cytokine measurements use $0.01\times$ PBS.

The differential measurement combines two layers of common-mode rejection. First, the reference wells experience the same buffer, temperature, and reference-electrode drift as the analyte wells, so
\begin{equation}
  \dIds^{\mathrm{sig}}(W_n,t) = \dIds(W_n,t) - \dIds(W_3,t),\quad n\in\{1,2\},
  \label{eq:differential}
\end{equation}
preserves binding-induced charge while suppressing drift. Second, within each well, cluster averaging across $\approx\pxwell{}$ pixels suppresses 1/f noise and threshold mismatch as \newline $\sigma_{\mathrm{eff}}/\sqrt{N_{\mathrm{cluster}}}$.

\section{Materials and Methods}
\label{sec:methods}

The experimental methods are organized according to the assay workflow. We first describe the universal surface-functionalization protocol that establishes a common amine-coupling chemistry across all sensing wells, followed by the allocation of capture antibodies and control chemistries within the four-well microfluidic architecture. We then detail the electrical characterization procedures used to establish the baseline sensor performance and the protocols used for cytokine dose-response, limit-of-detection, and specificity measurements.

\subsection{Materials}
\label{subsec:materials}
Capture antibodies were anti-human IL-6 (clone 5IL6, mouse IgG1; Thermo Fisher Scientific, M620)\cite{ThermoIL6_5IL6} and AF488-conjugated anti-human IL-10 (clone JES3-9D7, rat IgG1; Thermo Fisher Scientific, 53-7108-42).\cite{ThermoIL10_JES3} Goat anti-mouse IgG--AF488 (Thermo Fisher Scientific, A-11001) was used for fluorescence verification of W1. Recombinant human IL-6 (PeproTech, 200-06) and IL-10 (PeproTech, 200-10) were reconstituted to \SI{100}{\micro g/mL} in 0.1\% BSA in sterile water, aliquoted and stored at \SI{-80}{\celsius}. APTES (Thermo Scientific Pierce, 80370), 25\% glutaraldehyde (Thermo Scientific, A17876.AP), sodium cyanoborohydride (\nabhcn{}; Sigma-Aldrich, 156159), BSA ($\geq$98\%; Sigma-Aldrich, A7906), PBS ($1\times$, pH~7.4; Gibco, 10010031) and PBS-T (Pierce BupH, 28352) were used as received; anhydrous ethanol and sterile water were laboratory stock. \flexdym{} sheets were obtained from Eden Tech.\cite{EdenFlexDym}

\subsection{Universal Amine-Coupling Recipe}
\label{subsec:amine}

A common amine-coupling surface chemistry was used across all wells, with assay specificity determined solely by the immobilized capture antibody. The protocol was adapted from a previously validated \siN{} functionalization workflow used for Annexin~V immobilization in phosphatidylserine sensing\cite{NoyanPS2024} and follows established silane--aldehyde coupling procedures\cite{Vashist2014,Hermanson2013}.

The \siN{} surface was first cleaned using RCA-1 (\SI{70}{\celsius}, \SI{15}{min}) and activated by O$_2$ plasma (\SI{150}{W}, \SI{30}{s}) to generate surface hydroxyl groups. The surface was then silanized with 2\% APTES in anhydrous ethanol (\SI{90}{min}, room temperature) and cured (\SI{110}{\celsius}, \SI{15}{min}) to form a stable aminosilane layer. Primary amines were subsequently crosslinked using 2.5\% glutaraldehyde (\SI{50}{min}), followed by reduction with \SI{160}{mM} NaBH\textsubscript{3}CN (\SI{15}{min}) to stabilize the linkage.

Following surface preparation, the \flexdym{} overlay was bonded to the chip and capture antibodies were dispensed into the designated wells at \SI{10}{\micro g/mL} (\SI{60}{min}, room temperature, humidified environment). The wells were then rinsed with PBS-T, blocked with 1\% BSA (W1--W3), and washed prior to electrical measurements. Because the underlying chemistry is identical for all wells, different targets can be accommodated by substituting the capture probe while preserving the same surface-functionalization workflow. The complete protocol is provided in \suppref{S-I}.

\textit{Safety.}~RCA-1 (\ce{NH4OH}/\ce{H2O2}) and glutaraldehyde were handled in a fume hood with appropriate personal protective equipment. \nabhcn{} is toxic and releases hydrogen cyanide on contact with acid; it was handled only in a fume hood, kept away from acids, and disposed of as cyanide-containing waste according to institutional procedures. Oxygen-plasma treatment was performed in an enclosed plasma system.

\begin{figure*}[!t]
  \centering
  \includegraphics[width=\textwidth]{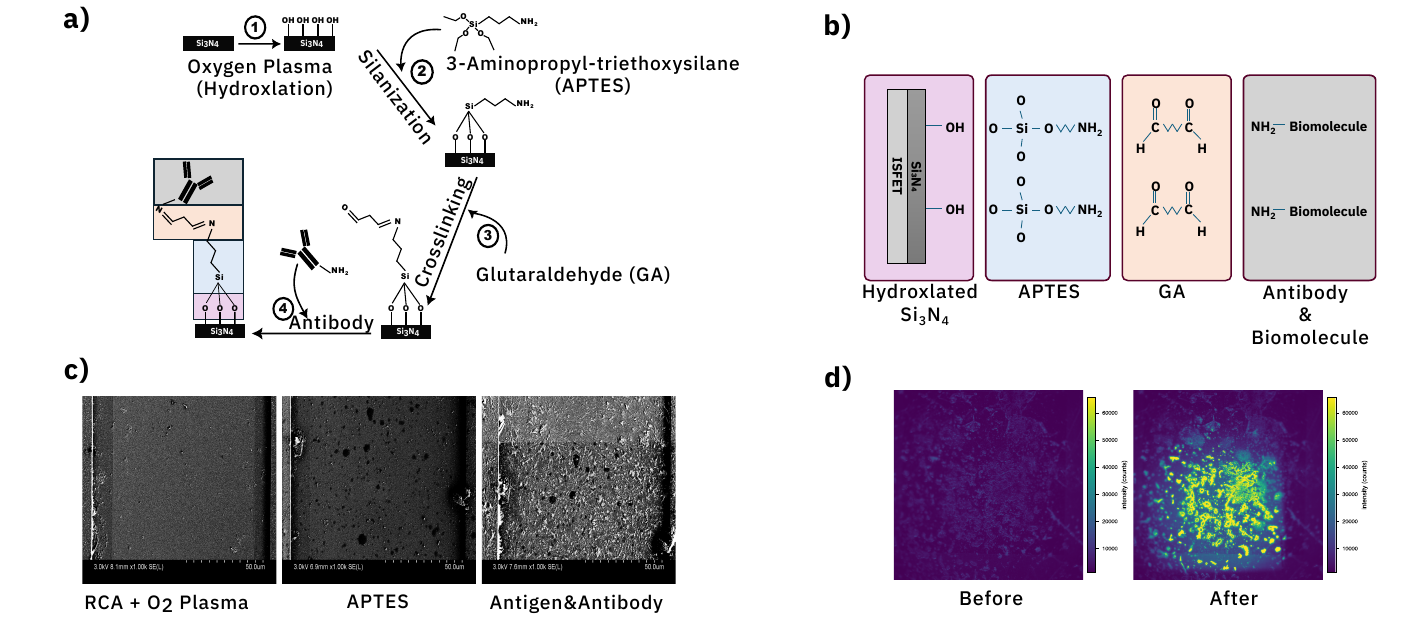}
  \caption{Universal amine-coupling functionalization of the \siN{} gate.
    \textbf{(a)}~Process flow: RCA $+$ O\textsubscript{2} plasma $\to$ 2\% APTES
    $\to$ \SI{110}{\celsius} cure $\to$ 2.5\% GA $\to$ \SI{160}{mM}
    NaBH\textsubscript{3}CN reduction $\to$ \SI{10}{\micro g/mL} capture antibody
    $\to$ 1\% BSA block. \textbf{(b)}~Corresponding amine-coupling chemistry.
    \textbf{(c)}~Step-wise SEM after RCA $+$ O\textsubscript{2} plasma, APTES,
    and antigen--antibody formation. \textbf{(d)}~AF488 fluorescence on the
    anti-IL-10 well (W2: JES3-9D7-AF488) before and after antibody incubation,
    confirming spatially resolved surface coverage.}
  \label{fig:chemistry}
\end{figure*}

\subsection{Four-Well Allocation}
\label{subsec:wells}
\begin{itemize}[leftmargin=*,nosep]
  \item \textbf{W1: anti-IL-6}---clone 5IL6 (mouse IgG1\cite{ThermoIL6_5IL6}); reports IL-6 via $\dIds(W_1)-\dIds(W_3)$.
  \item \textbf{W2: anti-IL-10}---clone JES3-9D7-AF488 (rat IgG1\cite{ThermoIL10_JES3}); reports IL-10. Direct AF488 verifies on-surface density.
  \item \textbf{W3: chemistry-only control}---APTES $+$ GA $+$ NaBH\textsubscript{3}CN, no antibody; quantifies non-specific binding to the aldehyde-functionalized surface and serves as the differential reference in Equation~(\ref{eq:differential}).
  \item \textbf{W4: specificity well}---antibody-functionalized and challenged with the off-target analyte to measure cross-reactivity; quantifies the specificity of the capture chemistry against the wrong cytokine.
\end{itemize}
The species mismatch between the capture antibodies (mouse 5IL6 on W1, rat JES3-9D7 on W2) is by design: a goat-anti-mouse-AF488 secondary fluoresces W1 without cross-binding W2.

\subsection{On-chip Electrical Characterization}
\label{subsec:methods_elec}
Chips are characterized in $1\times$ PBS at pH 4, 7, and 10. For each buffer, full $I_D$--$V_{GS}$ transfer curves are acquired on the central sensing region (a $4\times4$ pixel block at the array center) and full-array heatmaps are collected at fixed $V_{GS}$. Per-pixel pH sensitivity is extracted as the slope of $I_D$ versus pH; per-pixel $g_m$ at the operating point is the $I_D$--$V_{GS}$ derivative. A pH4$\to$pH10$\to$pH4 reversibility frame brackets the titration so drift can be separated from the pH response. Device-to-device variation is assessed by repeating this characterization on a second physical chip (Section~\ref{subsec:res_dev}). Per-cell transfer curves and the reversibility test are in \suppref{S-IV}.

\subsection{Dose-Response and Specificity Protocol}
\label{subsec:methods_dose}
After surface-functionalization validation, the chip was buffer-exchanged to $0.01\times$ PBS through at least three sequential washes to increase the Debye length for cytokine detection. Cytokine dose-response measurements were performed using a concentration ladder spanning \SI{100}{fM}--\SI{10}{nM} in $10\times$ increments. Each concentration (\SI{50}{\micro L} per well) was introduced sequentially from lowest to highest using \SI{1}{mL} and \SI{100}{\micro L} syringes, followed by a \SI{10}{min} incubation to allow binding and a \SI{1}{min} PBS-T rinse before the next injection. The primary measurement was the drain current, $I_D(t)$, acquired at a fixed gate bias of $V_{GS}=+2$~V from the central sensing pixels of each well. For each channel, $\Delta I_D$ was referenced to its pre-injection baseline, and the differential sensing signal was obtained by subtracting the response of the chemistry-only reference well (W3) according to (\ref{eq:differential}). The limit of detection (LOD) was defined as the lowest analyte concentration for which $|\Delta I_D^{\mathrm{sig}}|$ exceeded the measured $3\sigma$ non-specific detection floor (Section~\ref{subsec:res_dose}). The apparent $\mathrm{EC}_{50}$ (half-maximal effective concentration) was estimated as the concentration corresponding to half of the maximum measured response. Specificity was evaluated by challenging each functionalized well with the off-target cytokine (\SI{100}{nM}) and by exposing all wells to non-cytokine proteins (BSA, IgG, and lysozyme at \SI{1}{\micro M}) to quantify non-specific adsorption.
\section{Results}
\label{sec:results}
The results are presented in the order of the experimental validation workflow. Baseline electrical characterization first establishes the intrinsic performance of the CMOS \isfet{} array, including pH sensitivity, drift, noise, and transconductance, providing the reference operating conditions for biosensing measurements. The surface functionalization is then validated electrically, followed by characterization of IL-6 and IL-10 dose responses, limits of detection, and temporal kinetics. Finally, device-to-device reproducibility, multiplex operation, specificity, and \flexdym{} microfluidic integration are evaluated to assess the complete sensing platform.
\subsection{Baseline Electrical Characterization}
\label{subsec:res_baseline}

The array's baseline electrical behavior, per-pixel pH sensitivity (\qty{1.13}{\micro\ampere\per\text{pH}}), $1/f$-limited noise budget, and \qty{-8.9}{\micro\ampere\per\hour} drift are summarized in \suppref{S-IV}--S-V. This drift originates from ion migration and slow surface-potential relaxation at the
\siN{}--electrolyte interface, and is distinct from temperature-induced current changes\cite{NoyanDrift2024}; differential referencing against W3
(\ref{eq:differential}) removes it as a common-mode term. The operating-point transconductance $g_m$, is used to convert $\Delta I_{\text{DS}}$ to $\Delta V_{\text{th}}$.

\subsection{Surface-Chemistry Validation (on-chip)}
\begin{figure*}[!t]
  \centering
  \includegraphics[width=0.95\textwidth]{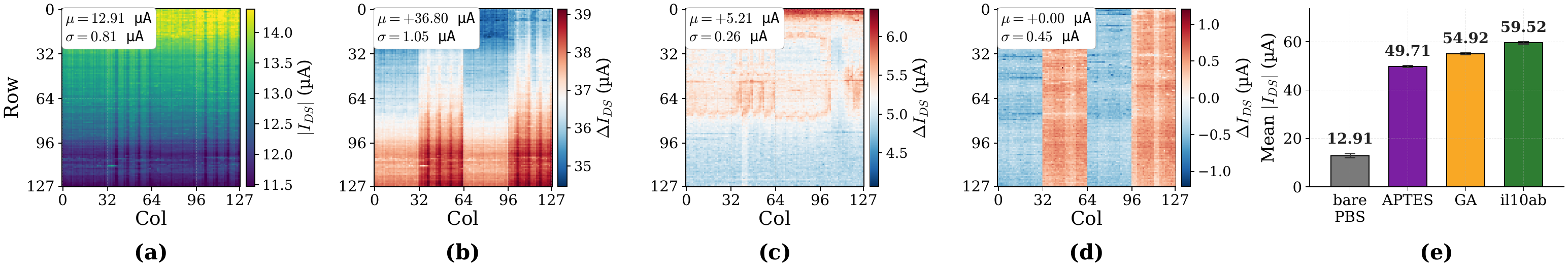}
\caption{Surface-chemistry progression on the full $128\times128$ array at a matched operating point. (a)~Bare chip in PBS. (b)~Silanization differential (APTES $-$ bare), mean $\dIds=+\SI{36.8}{\micro\ampere}$. (c)~Aldehyde crosslinking (GA $-$ APTES), $+\SI{5.21}{\micro\ampere}$. (d)~Capture-antibody spatial uniformity (mean-centered). (e)~Cumulative mean $|I_D|$ per step. The vertical gradient visible across the array reflects the row-by-row readout order, so it encodes slow baseline drift accumulated over the measurement period rather than true spatial non-uniformity. Each step is detectable above the per-step noise floor, so every chemistry step is electrically traceable on-chip.}
  \label{fig:chem_progress}
\end{figure*}

The chemistry steps each produce statistically resolved cumulative shifts (\autoref{fig:chem_progress}). Silanization produces the largest signal ($+\SI{36.8}{\micro\ampere}$), consistent with the positively charged --\ce{NH3+} APTES layer; GA adds $+\SI{5.21}{\micro\ampere}$, and the antibody layer a further $+\SI{4.6}{\micro\ampere}$. Per-pixel uniformity remains well controlled ($\sigma\leq\SI{1.05}{\micro\ampere}$), indicating uniform coating. Step-wise SEM and AF488 fluorescence (\autoref{fig:chemistry}c,d; \suppref{S-III}) provide orthogonal confirmation.

\subsection{Dose Response and Limit of Detection}
\label{subsec:res_dose}

The measured dose-responses for both cytokines are shown in \figdose, with the spatial (masked-well) rendering in Fig.~S6 (ESI\textsuperscript{\dag}). IL-10 rises steeply: the response is already a large fraction of its \SI{10}{nM} level at \SI{100}{fM}, indicating a high-affinity capture (apparent half-maximal $\mathrm{EC}_{50}\approx\ECilten{}$), and remains dose-graded across the full \SI{100}{fM}--\SI{10}{nM} window without saturating, reaching \platilten{} at \SI{10}{nM}. IL-6 is graded across the same window (apparent $\mathrm{EC}_{50}\approx\ECilsix{}$), reaching \platilsix{}. Reading each means-line down to the $3\sigma$ non-specific detection floor gives LODs of $\LODilten{}$ ($\LODiltenMass{}$) for IL-10 and $\approx\LODilsix{}$ ($\approx\LODilsixMass{}$) for IL-6, both at or below the healthy-serum baseline (\textless\SI{10}{pg/mL}\cite{Tanaka2014IL6}).

\textit{Basis of the detection limit.}~Ideal whole-well averaging predicts a sub-nanoampere noise floor, which would imply an unrealistically low limit of detection. In practice the limit is not set by thermal noise but by how reproducibly the baseline returns between injections (residual drift and re-equilibration) once non-specific binding is subtracted via the $W_3$ reference. We therefore report the LOD as the concentration at which the 
dose-response rises above this measured non-specific detection floor (the dotted line in \figdose), rather than the value implied by the averaged thermal floor. The non-specific detection floor is determined from three repeated measurements in each sample Well $W_1$ and $W_2$ and, as it provides a larger and more realistic estimate for limit of detection, we report it as the LOD for comparison against ELISA.

\subsubsection{Temporal kinetics}
\label{subsec:res_temporal}
The cumulative dosing trace (\figtemporal) resolves each concentration step within $\sim$\SI{30}{s} of injection and plateaus within $\sim$\SI{60}{s}, with plateau $\sigma\leq\SI{0.08}{\micro\ampere}$---well below the inter-step difference at clinically relevant concentrations. The near-saturation of the two highest steps is the temporal signature of the steep IL-10 affinity seen in \figdose.

\begin{figure*}[!t]
  \centering
  \includegraphics[width=\textwidth]{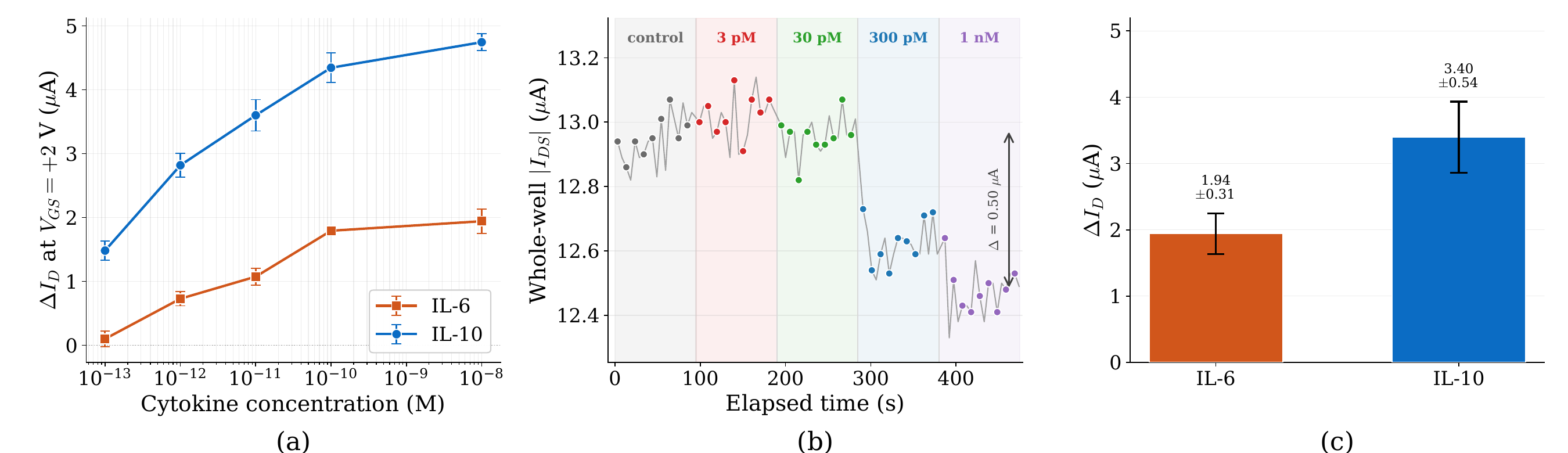}
  \caption{\textbf{(a)}~IL-6 and IL-10 dose-response ($\dIds$ at $V_{GS}=+2$~V,
  referenced to each cytokine's own pre-injection baseline) over
  \SI{100}{fM}--\SI{10}{nM}; lines join the per-concentration means, error bars
  are the across-pixel deviation. \textbf{(b)}~Time-domain IL-10 cumulative dose
  response on a $32\times32$ pixel cluster; shaded bands are the dose phases and
  the whole-well $|I_{DS}|$ steps monotonically over a
  \SI{0.82}{\micro\ampere} swing. \textbf{(c)}~Device-to-device variance on the
  IL-6 and IL-10 plateaus after per-device $g_m$ calibration
  ($\pm\devcalib{}$). Panels (a) and (c) report $|\dIds|$.}
  \label{fig:dose_curve}
\end{figure*}

\subsection{Device-to-Device Variation}
\label{subsec:res_dev}
Because prior single-device FET cytokine sensors leave device-to-device variability uncharacterized\cite{Hu2023IL6}, we quantify it directly across two physical chips. The base transfer characteristics differ markedly: saturation current by $\approx$ 16\% and peak $g_m$ by $\approx$ 62\% (\SI{14}{} vs \SI{36}{\micro\ampere/V}). Since the raw binding signal scales with transconductance ($\dIds=g_m\,\dVth$), the \emph{uncalibrated} device-to-device measurement variance is dominated by this $g_m$ spread, $\approx\devraw{}$. A one-time per-device $g_m$ calibration, reporting $\dVth=\dIds/g_m$ rather than raw $\dIds$, removes the gain term and leaves the saturation-current-level residual, $\approx\devcalib{}$ (\figdevvar). Translated to the analyte plateaus this is $\platilsix{}\pm\SI{0.47}{\micro\ampere}$ (IL-6) and $\platilten{}\pm\SI{0.86}{\micro\ampere}$ (IL-10) gm-calibrated, versus $\pm\SI{1.82}{}$ and $\pm\SI{3.35}{\micro\ampere}$ uncalibrated. The intra-die per-pixel transducer-gain CV ($\approx$13\%) is consistent with the calibrated device-to-device figure. We therefore report gm-calibrated $\dVth$ as the platform's primary metric, which keeps device-to-device variance near $\pm\devcalib{}$ rather than $\pm\devraw{}$.

\subsection{Multiplex Demonstration and Specificity}
\label{subsec:res_multiplex}
\begin{figure*}[!t]
  \centering
  \includegraphics[width=\textwidth]{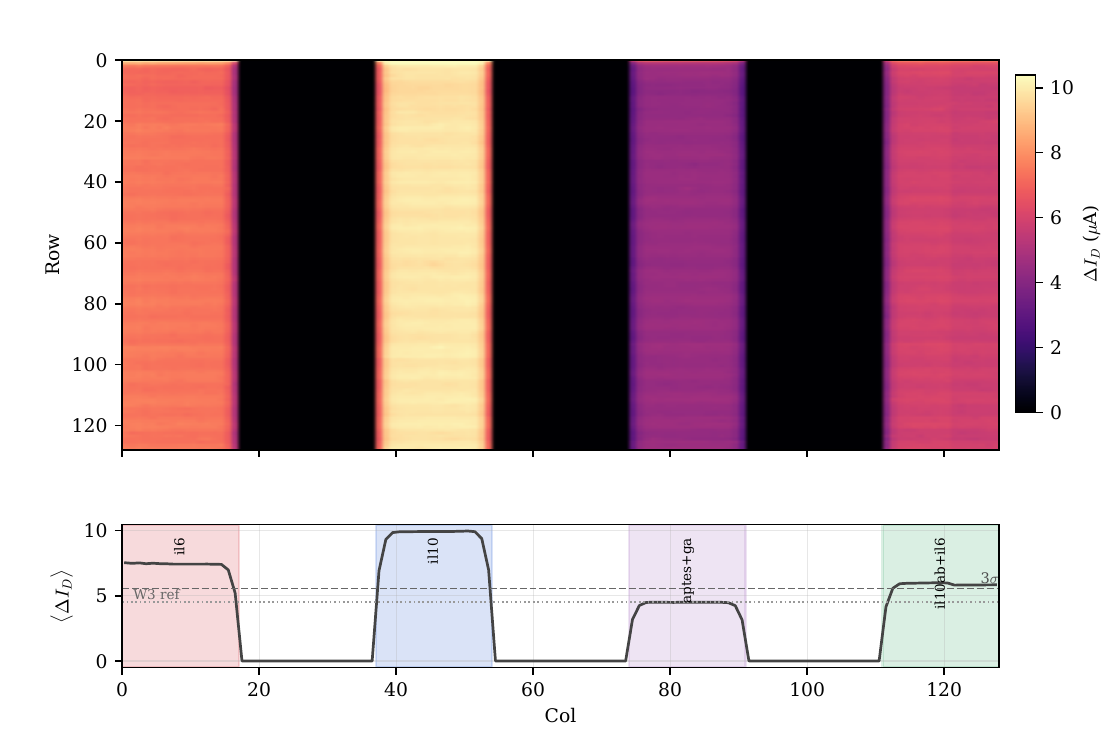}
  \caption{Measured four-well multiplex signature. $128\times128$ view and
column profile with IL-6 on the anti-IL-6 well (W1), IL-10 on the anti-IL-10 well (W2), the chemistry-only control (W3) at the non-specific floor, and the
specificity well (W4: anti-IL-10 challenged with off-target IL-6). Dashed line marks the $3\sigma$ non-specific detection floor.}
  \label{fig:multiplex_headline}
\end{figure*}

Four-well operation is shown in \autoref{fig:multiplex_headline}. With IL-6
dosed on the anti-IL-6 well (W1) and IL-10 on the anti-IL-10 well (W2), the two
channels respond independently at $\dIds \approx 7\mu A$ and $\approx 10 \mu A$,
respectively, while the chemistry-only reference (W3) stays at the non-specific
floor. Capture specificity is tested in W4, an anti-IL-10 well challenged with
IL-6: the response remains at $\approx 5.5 \mu A$, i.e. 7.1\% of the
on-target IL-6 signal in W1, below the 10\% multiplex-specificity criterion.
Wall regions read $\sim$0 throughout, confirming fluidic isolation.

Two elements of the specificity characterization remain. The reciprocal pairing
(IL-10 challenged onto an anti-IL-6 well) has not been run, so cross-reactivity is established in one direction only. 

\subsection{\flexdym{} Integration and Cross-Leak Performance}
\label{subsec:res_flexdym}
The \flexdym{} microfluidic overlay formed a stable, thermally bonded seal on the CMOS chip, preserving the functionalized sensing surface. Fluidic isolation between adjacent wells was verified by alternately introducing pH~4 and pH~10 buffer into neighboring channels and measuring the resulting threshold-voltage maps. Sharp transitions were observed at each well boundary, with a plateau-to-plateau difference of approximately \SI{121}{mV}, while the interiors of individual wells remained uniform within the measured per-well noise, indicating negligible cross-leakage over the assay duration. The reversible bond remained mechanically stable throughout all measurements. A quantitative analysis of well-to-well isolation, including boundary leakage, edge-transition width, and leakage as a fraction of the plateau response, is provided in \suppref{S-II}.

\section{Discussion}
\label{sec:discuss}

The platform addresses three persistent gaps in label-free CMOS cytokine sensing. First, dense-array cluster averaging converts the per-device sensitivity of individual SiNW-FETs\cite{Hu2023IL6,Rani2020} into a measurement qualifiable against ELISA, with both IL-6 and IL-10 single-analyte dose-responses measured and $3\sigma$ LODs at the healthy-serum baseline. Second, adopting \flexdym{} removes the absorption/leaching pathologies of PDMS\cite{Toepke2006,Regehr2009} while keeping an elastomeric, reversibly-bonded overlay. Third, allocating a chemistry-only control and a reference well makes the differential subtraction in (\ref{eq:differential}) defensible with built-in controls. The two-chip study further shows that the dominant reproducibility risk is transconductance spread ($g_m$ CV $\approx\devraw{}$) and that a per-device $g_m$ calibration is sufficient to suppress it to $\approx\devcalib{}$ -- a concrete, testable reporting recommendation that single-device studies cannot make.

Several limitations remain. (i)~Cross-specificity is characterized in one direction only: IL-6 was
challenged against an anti-IL-10 well (W4), but the reciprocal pairing
(IL-10 on anti-IL-6) and the non-cytokine interferents (BSA, IgG, lysozyme)
remain to be measured.(ii)~IL-10's steep low-end response (apparent $\mathrm{EC}_{50}\approx\ECilten{}$)
concentrates most of the dynamic range below $\sim$\SI{10}{pM}, so IL-10 is detectable but only weakly \emph{quantifiable} across the upper decades. (iii)~Antibody-mediated detection places the binding event $\sim$\SIrange{8}{15}{nm} from the gate, near the $0.01\times$ PBS Debye length, so absolute sensitivity is Debye-bounded; aptamer recognition\cite{Hu2023IL6,Ryazantsev2024} would reduce the binding distance and, in principle, gain an order of magnitude. (iv)~Validation here is in $1{:}100$-diluted buffer; undiluted serum will require anti-fouling chemistry, which we identify as future work.

The platform's universality is supported by its prior application to phosphatidylserine detection with Annexin~V chemistry\cite{NoyanPS2024}; the same APTES $+$ GA $+$ NaBH\textsubscript{3}CN backbone supports antibodies, lipid receptors, peptides, and amine-modified aptamers, suggesting a path to $8\times1$ panels (TNF-$\alpha$, IFN-$\gamma$, IL-1$\beta$, IL-8).

\section{Conclusion}
\label{sec:conclusion}

We have presented the architecture, surface chemistry, and validation of a CMOS \isfet{} pixel-array biosensor integrating a \flexdym{} thermoplastic-elastomer microfluidic partition for multiplexed label-free cytokine sensing. With the corrected $4\times1$ geometry (\num{17}-column wells, $\approx\pxwell{}$ pixels each, $\sqrt{N}\approx\rootN{}\times$ averaging), measured single-analyte dose-responses give IL-10 and IL-6 $3\sigma$ LODs of $\approx\LODilten{}$ ($\approx\LODiltenMass{}$) and $\approx\LODilsix{}$ ($\approx\LODilsixMass{}$) at the healthy-serum baseline, and a two-chip study shows that per-device $g_m$ calibration holds device-to-device measurement variance near $\pm\devcalib{}$. To our knowledge, this is the first CMOS \isfet{} biosensor to use SEBS rather than PDMS, the first to apply $\sqrt{N}$ cluster averaging to cytokine detection, and the first to quantify and calibrate device-to-device cytokine-measurement variance on a dense array.

\section*{Author contributions}
Utku Noyan: conceptualization, methodology, software, writing -- original draft. Sahil Shah: methodology, supervision, funding acquisition, writing -- review \& editing. Pamela Abshire: methodology, supervision, funding acquisition, writing -- review \& editing.

\section*{Conflicts of interest}
There are no conflicts to declare.

\section*{Ethical statement}
This study did not involve human participants, human samples or live animals; recombinant cytokines and commercial antibodies were used throughout.

\section*{Data availability}
 Raw array frames and analysis scripts are available from the corresponding authors upon reasonable request.

\section*{Acknowledgements}
The authors gratefully acknowledge the NIST-led Nanotechnology Xccelerator program and SkyWater Technology Foundry for providing the circuit design framework and the planarized \SI{130}{nm} CMOS fabrication platform that enabled the bioelectronic sensor integration in this work. We also thank Karina Torres Castro for her valuable technical discussions on surface chemistry and microfluidic integration, specifically for suggesting the use of Flexdym over PDMS to achieve more reliable thermal bonding and packaging.  Finally, we recognize Dr. Nihan Atak and Dr. Darwin R. Reyes from the National Institute of Standards and Technology (NIST), where the microfabrication of the Flexdym microfluidic device was conducted.

\bibliography{references_journal}

\clearpage
\onecolumn
\section*{Electronic Supplementary Information}
\noindent\rule{\textwidth}{0.4pt}

\setcounter{section}{0}
\setcounter{figure}{0}
\setcounter{table}{0}
\setcounter{equation}{0}
\renewcommand{\thesection}{S-\Roman{section}}
\renewcommand{\thesubsection}{S-\Roman{section}.\arabic{subsection}}
\renewcommand{\thefigure}{S\arabic{figure}}
\renewcommand{\thetable}{S\arabic{table}}
\renewcommand{\theequation}{S\arabic{equation}}


\section{Full Surface-Chemistry Protocol}
\label{supp:protocol}
Table~\ref{tab:s_protocol} gives the complete amine-coupling workflow. All steps are performed on bare die before \flexdym{} bonding except the antibody and block steps, which are dispensed per well after bonding.

\begin{table}[h]
\centering
\caption{Step-by-step surface-functionalization protocol.}
\label{tab:s_protocol}
\renewcommand{\arraystretch}{1.2}
\footnotesize
\begin{tabularx}{\textwidth}{@{}c p{3.0cm} p{3.2cm} p{2.0cm} p{1.6cm} X@{}}
\toprule
\textbf{\#} & \textbf{Step} & \textbf{Reagent / concentration} & \textbf{Time / temp} & \textbf{Volume} & \textbf{Notes / safety} \\
\midrule
1 & RCA-1 clean        & \ce{H2O}:\ce{NH4OH}:\ce{H2O2} (5:1:1) & \SI{15}{min}, \SI{70}{\celsius} & immerse & Fume hood; base-peroxide hazard \\
2 & DI rinse + dry     & \ce{H2O}, then N$_2$                  & ---                            & flood   & --- \\
3 & O$_2$ plasma       & \SI{150}{W}                     & \SI{30}{s}                     & ---     & March asher; proceed within \SI{2}{h} \\
4 & Silanization       & 2\% APTES in anhydrous EtOH           & \SI{90}{min}, RT               & immerse & Dry; APTES is moisture-sensitive \\
5 & EtOH rinse         & anhydrous EtOH ($\times3$)            & ---                            & flood   & --- \\
6 & Cure               & ---                                   & \SI{15}{min}, \SI{110}{\celsius}& ---     & Hotplate/oven \\
7 & Crosslink (GA)     & 2.5\% glutaraldehyde in PBS           & \SI{50}{min}, RT               & immerse & Fume hood; GA is a sensitizer \\
8 & Reduction          & \SI{160}{mM} NaBH\textsubscript{3}CN  & \SI{15}{min}, RT               & immerse & \textbf{Toxic}; HCN on acid contact---fume hood \\
9 & Capture antibody   & \SI{10}{\micro g/mL} in PBS (W1, W2, W4)& \SI{60}{min}, RT              & \SI{50}{\micro L} & Humidified; W1 5IL6, W2/W4 JES3-9D7-AF488 \\
10 & Wash              & PBS-T                                 & ---                            & flood   & --- \\
11 & Block             & 1\% BSA in PBS (all wells)            & \SI{30}{min}, RT               & \SI{50}{\micro L} & W3 is the BSA-only reference (no antibody) \\
12 & Wash              & PBS-T, then $0.01\times$ PBS          & ---                            & flood   & Buffer-exchange for low-Debye sensing \\
\bottomrule
\end{tabularx}
\end{table}

\section{Per-Pixel Sensitivity, Transfer Curves, and pH Reversibility}
\label{supp:reversibility}
The pixel-to-pixel sensitivity distribution (Fig.~\ref{fig:senstats}) gives a CV of 14\%, consistent with foundry-passivated \siN{} uniformity and with the intra-die transducer-gain spread used in the device-to-device analysis. Representative per-cell $I_D$--$V_{GS}$ transfer curves reproduce the subthreshold-to-saturation transition across cells, with the pH-induced lateral shift varying cell-to-cell at the level set by $\sigma_{\mathrm{px}}$---motivating cluster-averaged rather than single-pixel $\dVth$. A pH~4$\to$7$\to$10$\to$7$\to$4 cycle returns close to the starting response on the reverse leg, confirming reversibility with small hysteresis.

\begin{figure}[]
  \centering
  \includegraphics[width=\linewidth]{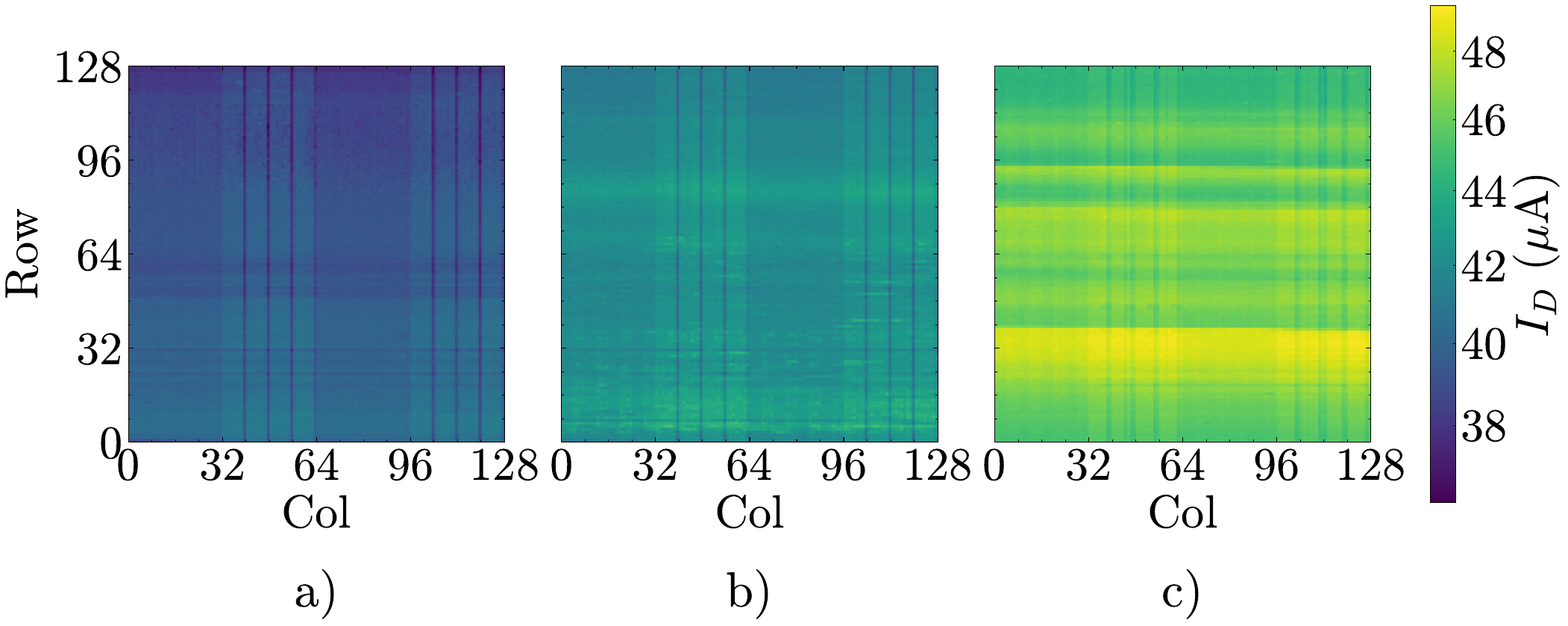}
  \caption{Spatial response of the $128\times128$ array at fixed $V_{GS}$ across pH 4, 7, and 10. The pH-dependent shift of mean \siN{} surface charge produces a monotonic shift of $|I_D|$ across the array. Row-banding and column-stripe artifacts reflect raster-scan drift and ADC offset; they are repeatable frame-to-frame and removed by differential referencing across consecutive scans.}
  \label{fig:fixedvgs}
\end{figure}

\begin{figure}[]
  \centering
  \includegraphics[width=\linewidth]{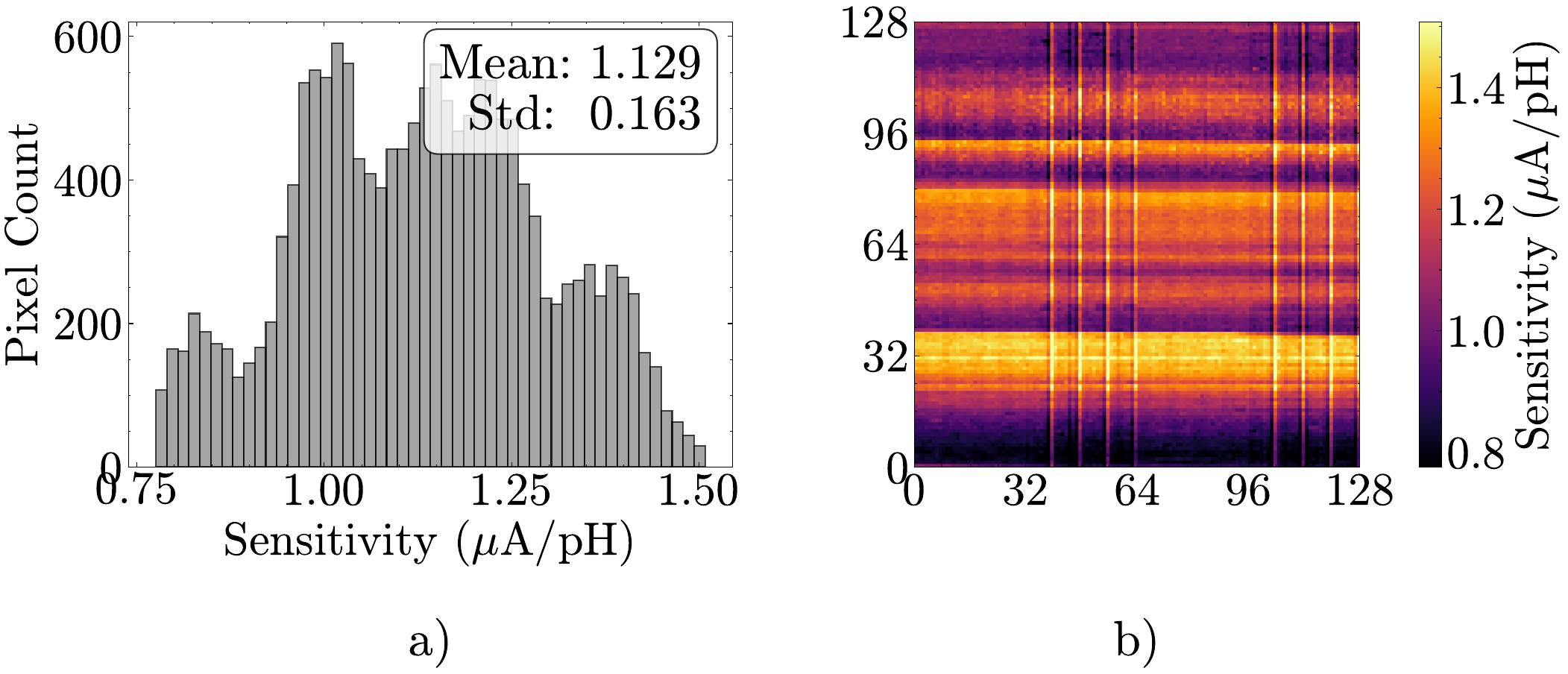}
  \caption{Per-pixel pH-sensitivity statistics on the array over the \num{16384} pixels: mean \SI{1.13}{\micro\ampere/pH}, $\sigma_{\mathrm{px}}=\SI{0.163}{\micro\ampere/pH}$ (CV~14\%). \emph{Left:} sensitivity histogram; \emph{right:} per-pixel sensitivity map, whose horizontal banding follows the row-direction raster artifact of \autoref{fig:fixedvgs}.}
  \label{fig:senstats}
\end{figure}
\begin{figure}[htbp]
  \centering
  \includegraphics[width=\linewidth]{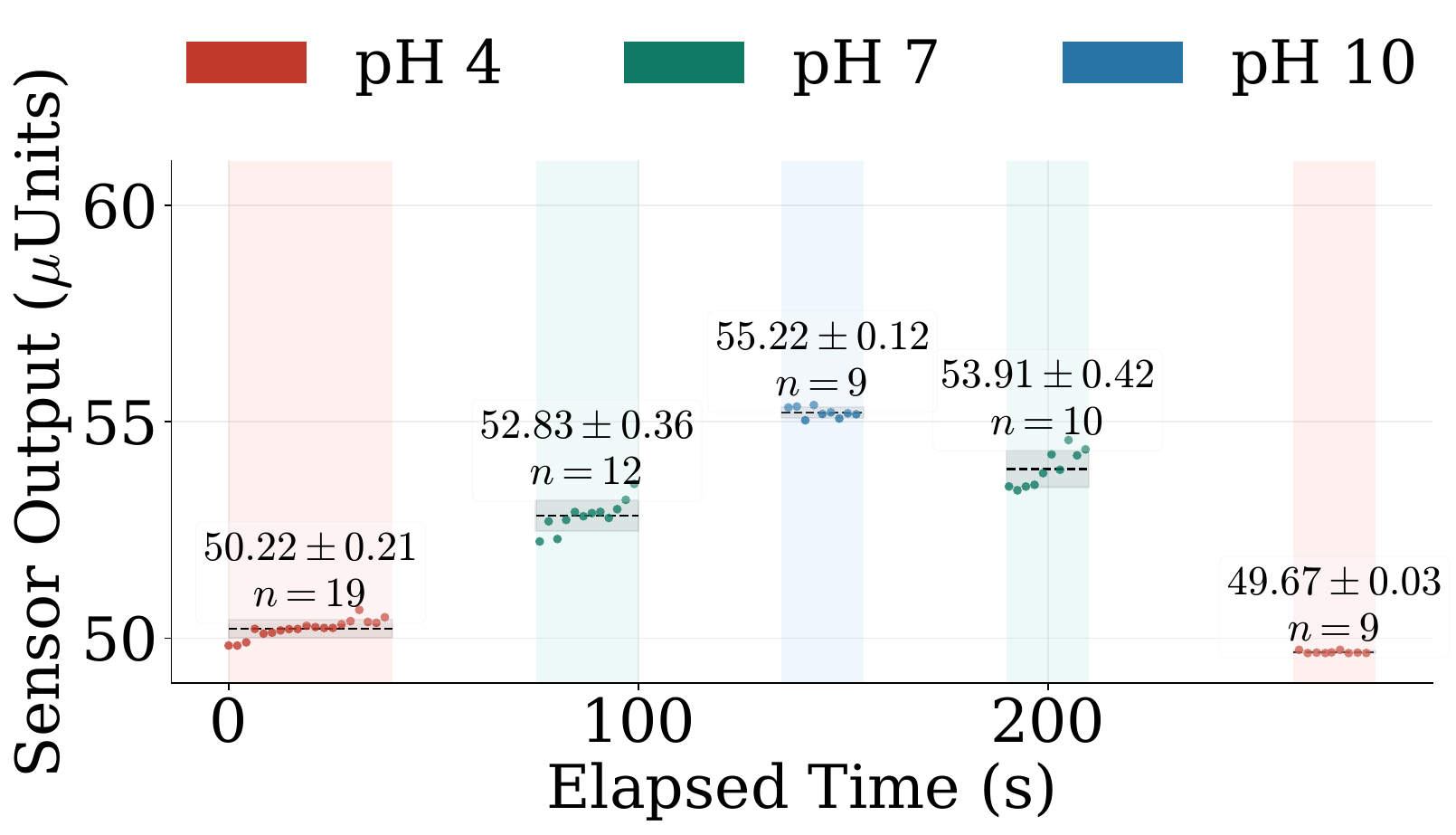}
  \caption{\textbf{Real-time ISFET sensor stability and repeatability across step changes in pH.} 
  Time-series sensor output ($\mu$Units) recorded across alternating buffer immersions ($\mathrm{pH}~4$, $\mathrm{pH}~7$, and $\mathrm{pH}~10$). The minimal deviation in $\mu$ across repeated immersions in identical buffer solutions demonstrates negligible sensor drift and high baseline reversibility.}
  \label{fig:stability_drift}
\end{figure}
\section{Noise, PSD, and Long-Duration Drift}
\label{supp:noise}
\begin{figure}[!t]
\centering\includegraphics[width=\linewidth]{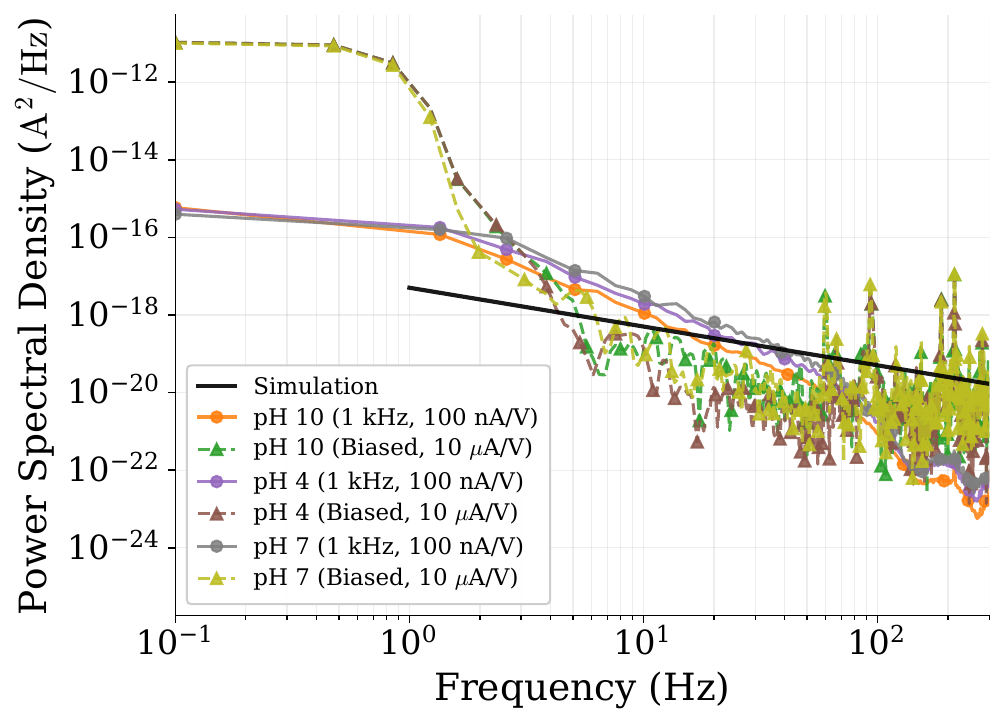}
\caption{Drain-current power spectral density of representative pixels at pH~4, 7, and 10, at two readout gains (\SI{100}{nA/V}, \SI{10000}{nA/V}). The biased traces show a $1/f$ corner near \SI{1}{Hz} and approach the simulated technology floor above \SI{10}{Hz}.}
\label{fig:s_psd}
\end{figure}

\begin{figure}[htbp]
  \centering
  \includegraphics[width=\linewidth]{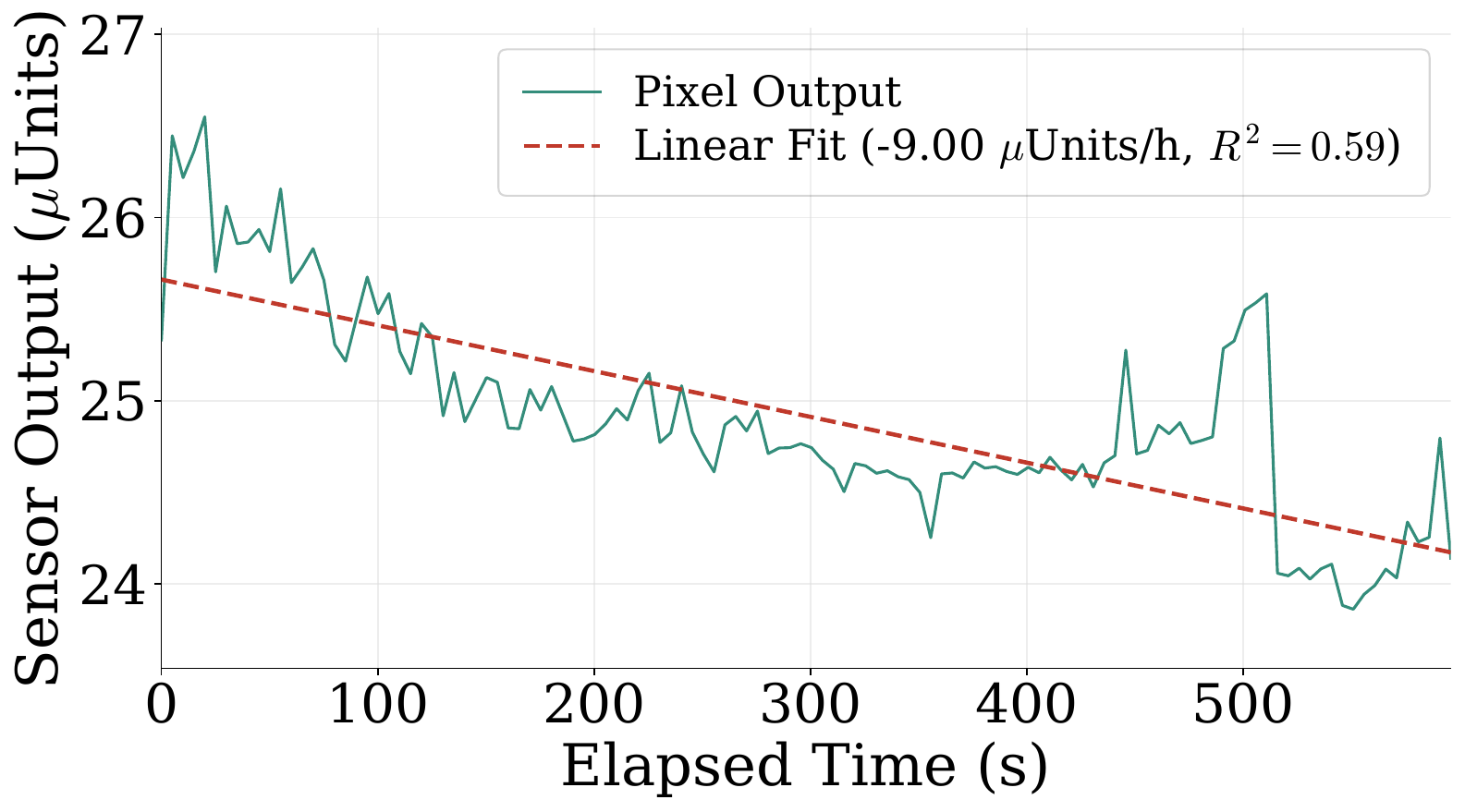}
\caption{\textbf{Continuous real-time ISFET baseline stability.} 
  Time-series output ($\mu$Units) recorded from pixel $\mathrm{R60C60}$ during extended immersion ($16.7~\text{h}$) in buffer solution. A first-order linear regression model (dashed red line) fit to the raw sensor output (solid green line) yields an exceptionally low baseline drift rate of $-0.05~\mu\text{Units/h}$ ($R^2 = 0.55$), demonstrating high interface stability and minimal ionic trapping within the passivation layer.}
  \label{fig:baseline_drift}
\end{figure}

The per-pixel input-referred noise is obtained by integrating the PSD (Fig.~\ref{fig:s_psd}) over the \SIrange{0.1}{10}{Hz} assay band and dividing by $g_m\approx\SI{22}{\micro\siemens}$: $\sqrt{\int S_I\,df}\approx\SI{0.445}{\micro\ampere}\Rightarrow\SI{20.2}{mV}$ per pixel. Cluster averaging over $N$ independent pixels scales the input-referred noise as $1/\sqrt{N}$: $N=\pxwell{}$ pixels (one well, $17\times128$) $\to$ \SI{0.43}{mV}. The detection statistic is the \emph{differential} signal $\dIds(W_n)-\dIds(W_3)$, whose blank scatter is $\sigma_{\mathrm{blank}}\approx\sqrt{2}\times\SI{0.43}{mV}\approx\SI{0.61}{mV}$, giving a $3\sigma_{\mathrm{blank}}$ threshold of \SI{1.84}{mV}. In practice the achievable LOD is set by this differential reproducibility (residual drift, re-equilibration; \autoref{fig:baseline_drift}) rather than the thermal floor, which is why the main-text LOD is quoted from the dose-response crossing of $3\sigma_{\mathrm{blank}}$.

\section{Temporal Binding Kinetics}
\label{supp:temporal}
The cumulative dosing trace (\figtemporal) resolves each concentration step within $\sim$\SI{30}{s} of injection and plateaus within $\sim$\SI{60}{s}, with plateau $\sigma\leq\SI{0.08}{\micro\ampere}$.

\section{Clinical Concentration Mapping}
\label{supp:clinical}
For IL-10 (MW $\approx$\SI{18.6}{kDa}), $1~\text{pM}=\SI{18.6}{pg/mL}$. Table~\ref{tab:s_clinmap} maps the dose ladder to clinical regimes; the quantitative window spans \SI{100}{fM}--\SI{10}{nM} and remains dose-graded without saturating, consistent with the main-text dose-response.
\begin{figure}[!t]
  \centering
  \includegraphics[width=\linewidth]{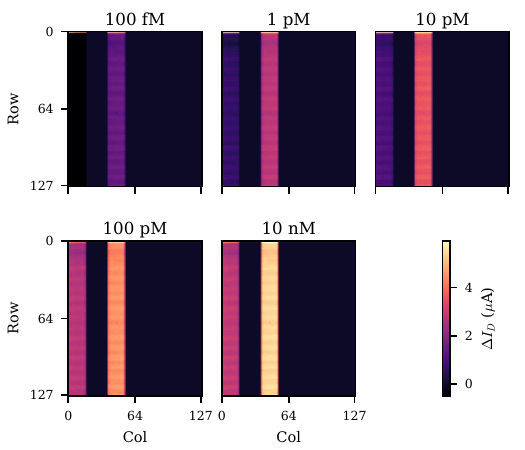}
  \caption{Dose response rendered as masked well heatmaps. The wells separate the measured IL-6 response in W1 and the IL-10 response in W2 on the full $128\times128$ array; columns are the concentration ladder (\SI{100}{fM}--\SI{10}{nM}). Walls and unaddressed wells read $\sim$0. Re-extracting each well through the core mask recovers the per-well dose values to within the per-pixel noise, validating the mask-based cluster averaging.}
  \label{fig:dose_spatial}
\end{figure}
\begin{table}[h]
\centering
\caption{IL-10 dose ladder mapped to mass concentration and clinical regime.}
\label{tab:s_clinmap}
\renewcommand{\arraystretch}{1.3}
\footnotesize
\begin{tabularx}{\columnwidth}{@{}l l X@{}}
\toprule
\textbf{Molar} & \textbf{Mass} & \textbf{Clinical regime} \\
\midrule
\textless\SI{100}{fM} (LOD) & \textless\SI{1.9}{pg/mL} & Healthy-serum baseline (\textless\SI{10}{pg/mL}) \\
\SI{1}{pM}   & \SI{18.6}{pg/mL}  & Low-grade inflammation \\
\SI{10}{pM}  & \SI{186}{pg/mL}   & Active inflammation \\
\SI{100}{pM} & \SI{1.86}{ng/mL}  & Sepsis / CRS range (\textgreater\SI{1}{ng/mL}) \\
\SI{1}{nM}   & \SI{18.6}{ng/mL}  & Severe CRS storm \\
\SI{10}{nM}  & \SI{186}{ng/mL}   & Top of tested window \\
\bottomrule
\end{tabularx}
\end{table}

\end{document}